\documentclass[11pt]{article}
\usepackage[ansinew]{inputenc}
\usepackage{graphicx,ae}
\usepackage{ae} 
\usepackage[T1]{fontenc}
\usepackage{amsmath,latexsym,amssymb}
\usepackage{color}
\usepackage[colorlinks]{hyperref}
\newcommand{\To}{\rightarrow}

\newcommand{\abs}[1]{\left\vert#1\right\vert}
\numberwithin{equation}{section}

\title{Generalized Non-extensive Statistical Distributions} 

\author{Oscar Sotolongo-Costa \\ {\it Department of Theoretical Physics. Havana University. Havana 10400, Cuba.}\\
\and Alejandro González González \\ {\it Department of Physics. University of Informatics Sciences, Havana, Cuba.}\\
\and Francois Brouers \\{\it Institute of Physics, Liège University, B5, 4000 Liège, Belgium.}}
\date{April 19, 2004}

\begin{document}
\maketitle

\begin{abstract}
We find the value of constants related to constraints in characterization of some known statistical distributions and then we proceed to use the idea behind maximum entropy principle to derive generalized version of this distributions using the Tsallis' and Renyi information measures instead of the well-known Bolztmann-Gibbs-Shannon. These generalized distributions will depend on $q \in \mathbb{R}$ and in the limit $q \To 1$ we obtain the ``classical" ones. We found that apart from a constant, generalized versions of statistical distributions following Tsallis' or Renyi are undistinguishable.
\end{abstract}
\bigskip
\section*{Introduction}
The development of statistical mechanics and its application to a wide variety of physical phenomena take a leap forward with the works of Shannon \cite{Shannon} and Jaynes \cite{Jay1,Jay2}. Before that, the usual line of reasoning was to construct the theory based on the equations of motion, supplemented by additional hypotheses of ergodicity, metric transitivity, or equal {\it a priori} probabilities and the identification of entropy was made at the end by comparison of the resulting equations with the laws of thermodynamics.\\\\
The point of view developed by Shannon and Jaynes took the concept of entropy to play a central role. In this modern approach, the fact that a probability distribution maximizes the entropy subject to certain constraints becomes essential to the justification of the use of that distribution for statistical inference. By freeing the theory from apparent dependence of physical hypotheses they made its principles and mathematical methods to become available for the treatment of many new physical problems. The constraints used to maximize the entropy are interpreted as the real unbiased knowledge about the system (as the mean energy, mean number of particles, etc) with the additional restriction of normalization of the probability distribution over the entire value range of the state variable. This is sometimes referred as the {\it Maximum Entropy Principle} (MEP)\cite{Jay1,Jay2}.\\\\ 
The Boltzmann-Gibbs entropy, usually referred as the Boltzmann-Gibbs-Shannon  entropy (we only write its continuous form): 
\begin{eqnarray}
S_{BGS}=-k\int p(x)\ln p(x)dx
\end{eqnarray}
has been at the basis of invaluable scientific results in physics, chemistry and elsewhere. In particular, the results are very well known for the treatment of equilibrium thermodynamics in systems composed of many independent subsystems where the entropy of individual independent parts are additive ad the entire systems has ergodic stationary states. \\\\
In order to handle with non equilibrium states in systems with a complex behavior (typically nonergodic), C. Tsallis proposed the entropic functional:
\begin{eqnarray}
S_{q}=k\frac{1}{q-1}\left[1-\int_{I}p^{q}(x)dx \right] \quad q\in \mathbb{R}
\end{eqnarray}
which is, in some sense, a generalization of the former entropic functional because in the limit for the Tsallis entropy $q\rightarrow 1$, we obtain again the Boltzmann-Gibbs-Shannon (BGS) entropy. The index $q$ has been interpreted as a degree of non extensivity which accounts for the case of many non independent or long-range interacting systems \cite{Tsallis1}.\\\\
The probability distributions obtained by maximization of this entropic functional with appropriate constraints with $q\neq1$ yields a correct description for different classes of systems: those with long range interactions and those with fluctuations of temperature or of energy dissipation. Examples cover a wide range of applications from hydrodynamic turbulence \cite{Arimitsu,Beck1,Beck2}, scattering processes in particle physics \cite{Bediaga}, self-gravitating systems is astrophysics \cite{Plastino} to  anomalous diffusing micro-organisms \cite{Upadhyaya}, classical and quantum chaos \cite{Weinstein,Lyra}, economics \cite{Borland} and others \cite{Tsallis1,Tsallis2}. Tsallis' formalism seems to be more general than BGS and there are reasons to believe that the $S_{q}$ is the proper physical entropy for generalizing the BGS formalism. In any case the BGS entropic form is obtained from the Tsallis one in the limit $q\rightarrow 1$. There are still other entropic forms like the Renyi one which has the property of being additive for independent systems.\\\\
It is well known that probability distributions plays a central role in statistical theories because they are the mathematical models of certain phenomena under analysis. Known distributions such as Normal, Log-Normal, Exponential, or Cauchy can be derived by maximization of the BGS entropy with appropriate constraints along with the usual normalization condition $\int_{I}p(x)dx=1$ where $I$ vary accordingly and $x$ is a continuous random variable with distribution $p(x)$. The constraints are identified with moments of order $n\geq1$ of the distribution.    \cite{Gokhale,Kagan,Kapur1,Kapur2}. \\\\
So it will be interesting to see if using the same idea behind the maximization of the entropy we can obtain generalized versions of some of known probability densities by replacing both the entropic functional and the constraints, chosen in a convenient manner. This paper revisit the idea in \cite{Gokhale} and we find the values of undetermined constants that appear related to constraints with BGS formalism. Then we find the generalized versions of the same statistical distributions by replacing the BGS entropic functional with Tsallis' and Renyi measures. In cases where logarithmic and exponential functions appear, they are replaced with its generalized versions as defined in \cite{Kania,Naudts}.

\section{Characterization of continuous ``classical" distributions} 

Maximum entropy principle states that the least-biased description that can be devised on the basis of some specific data is that which maximizes the thermodynamical or BGS entropy \cite{Jay1,Jay2}. Using this idea, characterization of statistical distributions via MEP is obtained when one maximizes the BGS entropy with some constraints (which not necessarily represent observable quantities) along with normalization condition. So, the statement of the problem is: ``let X be a continuous random variable with $x\in I$ and probability density $p(x)$ on $I$. Let the entropy be given by:
\begin{eqnarray}  
S=-\int_{I}p(x)\ln p(x)dx \qquad (BGS)
\end{eqnarray}  
where $p(x)$ is consistent with $N$ constraints conditions $f_{k}\left(x,p(x)\right), k=1 \ldots N$ and with the normalization condition on $I$, $\int_{I}p(x)dx=1$. Then the least-biased $p(x)$ is that which maximizes $S^{*}$:
\begin{eqnarray}  
S^{*}=S-\lambda_{o}\int_{I}p(x)dx-\sum_{k=1}^{N}\lambda_{k}f_{k}(x,p(x))
\end{eqnarray}  
where ${\lambda_{k}}$ are the Lagrange multipliers".

\subsection{Uniform Distribution: $I=(0,1)$} 
In this case the only constraint is the normalization condition, then we need to maximize:
\begin{eqnarray}
\nonumber S^{*}=S-\lambda_{0}\int_{0}^{1}p(x)dx \qquad giving: \\
\nonumber p(x)=e^{-(1+\lambda)}=\frac{1}{A}
\end{eqnarray}
For any bounded subinterval of $(0,1)$, ($(a,b)\subset(0,1)$) where $p(x)$ is given we have:
\begin{equation} 
\nonumber \int_{0}^{1}p(x)dx=\int_{a}^{b}p(x)dx=\frac{1}{A}(b-a)=1 \qquad \therefore A=b-a
\end{equation}
Then the uniform distribution is :
\begin{equation} 
p(x)=\frac{1}{b-a}; \qquad \forall x \in I
\end{equation}

\subsection{Exponential Distribution: $I=(0,\infty)$} 
The constraints are:\\\\
$1.\; \int_{0}^{\infty}p(x)dx=1 \qquad 2.\;\int_{0}^{\infty}xp(x)dx=m$\\\\
Then we need to maximize:
\begin{eqnarray}
\nonumber S^{*}=S-\lambda_{0}\int_{0}^{\infty}p(x)dx-\lambda_{1}\int_{0}^{\infty}xp(x)dx, \qquad giving:\\
\nonumber p(x)=C_{0}e^{-\lambda_{1}x}\qquad where \quad C_{0}=e^{-(1+\lambda_{0})}
\end{eqnarray}
Applying condition 1 we obtain $C_{0}=\lambda_{1}$ with $\lambda_{1}>0$ and applying condition 2 we obtain:
\begin{equation}
\nonumber \int_{0}^{\infty}xp(x)dx=\frac{1}{\lambda_{1}}=m 
\end{equation}
Then the exponential distribution is:
\begin{equation}
p(x)=me^{-mx}; \quad m>0 \qquad \forall x\in I
\end{equation}
with $m$ being the mean value of $x$ under the exponential distribution.

\subsection{Gamma Distribution: $I=(0,\infty)$} 
The constraints are:\\\\
$1.\;\int_{0}^{\infty}p(x)dx=1 \qquad  2.\;\int_{0}^{\infty}xp(x)dx=g_{1} \qquad 3.\;\int_{0}^{\infty}\ln(x)p(x)dx=g_{2}$\\\\
Then we need to maximize:
\begin{eqnarray}
\nonumber S^{*}=S-\lambda_{0}\int_{0}^{\infty}p(x)dx-\lambda_{1}\int_{0}^{\infty}xp(x)dx-
\lambda_{2}\int_{0}^{\infty}\ln(x)p(x)dx \\
\nonumber then \quad p(x)=C_{0}e^{-\lambda_{1}x}x^{-\lambda_{2}} \qquad where \quad C_{0}=e^{-(1+\lambda_{0})}
\end{eqnarray}
Applying condition 1 we get $\frac{C_{0}\Gamma(w)}{\lambda_{1}^{w}}=1$  with $w=1-\lambda_{2}>0$. Applying condition 2 and 3 we get $g_{1}=\frac{w}{\lambda_{1}}$ and $g_{2}=\frac{\Gamma^{'}(w)}{\Gamma(w)}-\ln\lambda_{1}$. So, the Gamma distribution is:
\begin{eqnarray}
p(x)=\frac{\lambda_{1}(\lambda_{1}x)^{w-1}e^{-\lambda_{1}x}}{\Gamma(w)}; \quad w>0, \quad \forall x\in I
\label{Gamma}
\end{eqnarray}

\subsection{Log-Normal Distribution: $I=(0,\infty)$} 
The constraints are: \\\\
$1.\;\int_{0}^{\infty}p(x)dx=1 \qquad 2.\;\int_{0}^{\infty}\ln(x)p(x)dx=0\qquad 3.\;\int_{0}^{\infty}\ln^{2}(x)p(x)dx=\sigma^{2}$\\\\
And we need to maximize:
\begin{eqnarray}
\nonumber S^{*}=S-\lambda_{0}\int_{0}^{\infty}p(x)dx-\lambda_{1}\int_{0}^{\infty}\ln(x)p(x)dx-
\lambda_{2}\int_{0}^{\infty}\ln^{2}(x)p(x)dx \\
\nonumber then \quad p(x)=C_{0}x^{-\lambda_{1}}e^{-\lambda_{2}\ln^{2}(x)} \qquad where \quad C_{0}=e^{-(1+\lambda_{0})}
\end{eqnarray}
Assuming $\lambda_{1}=1$ and applying condition 1 we get $C_{o}=\sqrt{\lambda_{2}/\pi}$ and applying condition 2 and 3 we get $\lambda_{2}=1/2\sigma^{2}$. So, the Log-Normal distribution is:
\begin{eqnarray}
p(x)=\frac{1}{\sigma x\sqrt{2\pi}}\exp\left(-\frac{\ln^{2}x}{2\sigma^{2}}\right)\qquad \forall x\in I
\end{eqnarray}

\subsection{Normal Distribution: $I=(-\infty,\infty)$} 
The constraints are: \\\\
$1.\;\int_{-\infty}^{\infty}p(x)dx=1 \qquad 2.\;\int_{-\infty}^{\infty}xp(x)dx=0\qquad 3.\;\int_{-\infty}^{\infty}x^{2}p(x)dx=\sigma^{2}$\\\\
We need to maximize:
\begin{eqnarray}
\nonumber S^{*}=S-\lambda_{0}\int_{-\infty}^{\infty}p(x)dx-\lambda_{1}\int_{-\infty}^{\infty}xp(x)dx-
\lambda_{2}\int_{-\infty}^{\infty}x^{2}p(x)dx \\
\nonumber then \quad p(x)=C_{0}e^{-\lambda_{1}x}e^{-\lambda_{2}x^{2}} \qquad where \quad C_{0}=e^{-(1+\lambda_{0})}
\end{eqnarray}
Assuming $\lambda_{1}=0$ and applying condition 1 we get $C_{o}=\sqrt{\lambda_{2}/\pi}$ and applying condition 2 and 3 we get $\lambda_{2}=1/2\sigma^{2}$. So, the Normal distribution is:
\begin{eqnarray}
p(x)=\frac{1}{\sigma\sqrt{2\pi}}\exp\left(-\frac{x^{2}}{2\sigma^{2}}\right)\qquad \forall x\in I
\end{eqnarray}

\subsection{Laplace Distribution: $I=(-\infty,\infty)$} 
The constraints are: \\\\
$1.\;\int_{-\infty}^{\infty}p(x)dx=1\qquad 2.\;\int_{-\infty}^{\infty}\abs x p(x)dx=w$\\\\
We need to maximize:
\begin{eqnarray}
\nonumber S^{*}=S-\lambda_{0}\int_{-\infty}^{\infty}p(x)dx-\lambda_{1}\int_{-\infty}^{\infty}\abs x p(x)dx\\
\nonumber p(x)=C_{0}e^{-\lambda_{1}\abs x} \qquad where \quad C_{0}=e^{-(1+\lambda_{0})}
\end{eqnarray}
Applying condition 1 we obtain: $C_{0}=\lambda_{1}/2$ and applying condition 2 we get: $w=1/\lambda_{1}$, so the Laplace distribution is:
\begin{eqnarray}
p(x)=\frac{\lambda_{1}}{2}e^{-\lambda_{1}\abs x}\qquad with \qquad \lambda_{1}\in \mathbb{R^{+}}
\end{eqnarray}


\section{Characterization of continuous ``generalized" distributions} 
In this section we derive generalized statistical distributions by replacing the BGS entropy measure with Tsallis and Renyi measures:
\begin{eqnarray}
S_{q}=\frac{1}{q-1}\left[1-\int_{I}p^{q}(x)dx \right]\qquad (Tsallis)\\
S_{R}=\frac{1}{1-q}\ln\left[\int_{I}p^{q}(x)dx\right] \qquad (Renyi)
\end{eqnarray}
and we use generalized logarithms \cite{Naudts} instead of natural logarithms. The generalized statistical distributions will give, in the limit $q\To 1$, the corresponding ``classical" ones, in a same way as the Tsallis and Renyi entropies give the BGS entropy when $q\To 1$ so, the statement of the problem remains the same. \\
In order to obtain the generalized distributions it is necessary to apply partial derivatives to Tsallis and Renyi measures. This derivatives have the same structure, the only difference being a constant factor appearing in Renyi derivative, namely $\int_{I}p^{q}(x)dx$:
\begin{eqnarray}
\nonumber \frac{\partial S_{q}}{\partial p(x)}=-\frac{q}{q-1}\int_{I}p^{q-1}(x)dx \\
\nonumber \frac{\partial S_{R}}{\partial p(x)}=\left(\frac{q}{1-q}\right)\frac{1}{A}\int_{I}p^{q-1}(x)dx \qquad with \qquad A=\int_{I}p^{q}(x)dx
\end{eqnarray}
So, in what follows, we use {\it only} the Tsallis measure to obtain the generalized densities $p(x)$ and the ``escort" probabilities defined as: 
\begin{equation}
P_{esc}(x)=\frac{p^{q}(x)}{\int p^{q}(x)dx}
\end{equation}

\subsection{Generalized Uniform Distribution: $I=(0,1)$} 
In this case the only constraint is the normalization condition, then we need to maximize:
\begin{eqnarray}
\nonumber S^{*}_{q}=S_{q}-\lambda_{0}\int_{0}^{1}p^{q}(x)dx \qquad giving: \\
\nonumber p(x)=\left[\frac{\lambda_{0}(1-q)}{q}\right]^{\frac{1}{q-1}}=C
\end{eqnarray}
For any bounded subinterval of $(0,1)$, ($(a,b)\subset(0,1)$) where $p(x)$ is given we have:
\begin{equation} 
\nonumber \int_{a}^{b}p^{q}(x)dx=C^{q}\int_{a}^{b}dx=C^{q}(b-a)
\end{equation}
Then the uniform distribution and the escort probabilities are:
\begin{eqnarray} 
p(x)=\left[\frac{\lambda_{0}(1-q)}{q}\right]^{\frac{1}{q-1}}=C\\
P_{esc}(x)=\frac{p^{q}(x)}{\int p^{q}(x)dx}=\frac{C^{q}}{C^{q}(b-a)}=\frac{1}{b-a}\qquad \forall x \in I=(0,1)
\end{eqnarray}

\subsection{Generalized Exponential Distribution: $I=(0,\infty)$} 
The constraints are:\\\\
$1.\; \int_{0}^{\infty}p(x)dx=1 \qquad 2.\;\int_{0}^{\infty}xp^{q}(x)dx=m$\\\\
Then we need to maximize:
\begin{eqnarray}
\nonumber S^{*}_{q}=S_{q}-\lambda_{0}\int_{0}^{\infty}p(x)dx-\lambda_{1}\int_{0}^{\infty}xp^{q}(x)dx, \qquad giving:\\
\nonumber p(x)=C[1 + (q-1)\lambda_{1}x]^{-\frac{1}{q-1}}  \qquad where \quad C=\left[\frac{\lambda_{0}(1-q)}{q}\right]^{\frac{1}{q-1}}
\end{eqnarray}
Applying condition 1 we obtain:
\begin{eqnarray}
\nonumber 1=\lim_{\xi\To\infty}\frac{C}{\lambda_{1}(q-2)}\left[ \xi ^{{\frac {q-2}{q-1}}}-1 \right]=\frac{C}{\lambda_{1}(2-q)}
\end{eqnarray}
which converges only if $1<q<2$ for $q\in\mathbb{R}$\\
Applying condition 2 we obtain:
\begin{eqnarray}
\nonumber m=\int_{0}^{\infty}xp^{q}(x)dx=C\int_{0}^{\infty}\frac{xdx}{[1 + (q-1)\lambda_{1}x]^{\frac{q}{q-1}}}=\frac{1}{\lambda_{1}}
\end{eqnarray}
Then the exponential distribution and the escort probability are:
\begin{eqnarray}
p(x)=\frac{\lambda_{1}(2-q)}{\left[1+\lambda_{1}(q-1)x\right]^{\frac{1}{q-1}}}\qquad \forall x \in I=(0,\infty),\quad q\in (1,2)\\
P_{esc}(x)=\frac{p^{q}(x)}{\int p^{q}(x)dx}=
\frac{\lambda_{1}^{q}(2-q)^{q-1}}{\left[1+\lambda_{1}(q-1)x\right]^{\frac{q}{q-1}}}
\end{eqnarray}
Is easy to verify that taking limit in $P_{esc}$ we get:
\begin{eqnarray}
\nonumber \lim_{q\To 1}\frac{\lambda_{1}^{q}(2-q)^{q-1}}{\left[1+\lambda_{1}(q-1)x\right]^{\frac{q}{q-1}}}=
\lambda_{1}e^{-\lambda_{1}x}
\end{eqnarray}
which is the exponential distribution.

\subsection{Generalized Gamma Distribution: $I=(0,\infty)$} 
The constraints are:\\\\
$1.\;\int_{0}^{\infty}p(x)dx=1 \qquad  2.\;\int_{0}^{\infty}xp^{q}(x)dx=g_{1} \qquad 3.\;\int_{0}^{\infty}ln_{q}(x)p^{q}(x)dx=g_{2}$\\\\
Then we need to maximize:
\begin{eqnarray}
\nonumber S^{*}_{q}=S_{q}-\lambda_{0}\int_{0}^{\infty}p(x)dx-\lambda_{1}\int_{0}^{\infty}xp^{q}(x)dx-
\lambda_{2}\int_{0}^{\infty}\ln_{q}(x)p^{q}(x)dx \\
\nonumber then \qquad p(x)=C[(q-1)\lambda_{1}x+x^{1-q}]^{-\frac{1}{q-1}}  \qquad where \quad C=\left[\frac{\lambda_{0}(1-q)}{q}\right]^{\frac{1}{q-1}}
\end{eqnarray}
with $\lambda_{2}=1$. Assuming that in the final expression for $p(x)$ the limit exists, we will take the limit inside the integral when applying the conditions:
\begin{eqnarray}
\nonumber 1=\lim_{q\To 1}\int_{0}^{\infty}p(x)dx=\int_{0}^{\infty}\lim_{q\To 1}p(x)dx=\frac{C}{\lambda_{1}^{2}}
\end{eqnarray}
Applying the second condition and using the last limit we get:
\begin{eqnarray}
\nonumber \int_{0}^{\infty}\lim_{q\To 1}\left[ xp^{q}(x)\right]dx=\frac{2}{\lambda_{1}}=g_{1}
\end{eqnarray}
The third condition gives: 
\begin{eqnarray}
\lim_{q\To 1}\int_{0}^{\infty}ln_{q}(x)p^{q}(x)dx=\int_{0}^{\infty}\lim_{q\To 1}p^{q}(x)\left[\frac{x^{1-q}-1}{1-q}\right]dx\\
\nonumber C\int_{0}^{\infty}x\ln (x)e^{-\lambda_{1}x}dx=C\left[\frac{1-\gamma-\ln(\lambda_{1})}{\lambda_{1}^{2}}\right]=
1-\gamma-\ln(\lambda_{1})=g_{2}
\end{eqnarray}
where $\gamma$ is the constant of Euler-Mascheroni defined as:
\begin{equation}
\nonumber \gamma=-\int_{0}^{\infty}e^{-u}\ln(u)du=-\Gamma'(1)
\end{equation} 
Then the generalized Gamma distribution and the escort probabilities are the same:
\begin{eqnarray}
P_{esc}(x)=\frac{p^{q}(x)}{\int_{0}^{\infty}p^{q}(x)dx}=\lambda_{1}^{2q}\left[\lambda_{1}x(q-1)+
x^{1-q}\right]^{\frac{q}{1-q}}\\
\nonumber \qquad \forall x \in I=(0,\infty),\quad q\in(1,2) 
\end{eqnarray}
Taking limit in this last expression we get the Gamma distribution for $w=2$ (see \ref{Gamma}):
\begin{equation}
\nonumber p(x)=x\lambda_{1}^{2}e^{-\lambda_{1}x}
\end{equation}

\subsection{Generalized Log-Normal Distribution: $I=(0,\infty)$} 
The constraints are: \\\\
$1.\;\int_{0}^{\infty}p(x)dx=1 \qquad 2.\;\int_{0}^{\infty}ln_{q}(x)p^{q}(x)dx=0\qquad 3.\;\int_{0}^{\infty}ln_{q}^{2}(x)p^{q}(x)dx=\sigma^{2}$\\\\
And we need to maximize:
\begin{eqnarray}
\nonumber S^{*}_{q}=S_{q}-\lambda_{0}\int_{0}^{\infty}p(x)dx-\lambda_{1}\int_{0}^{\infty}ln_{q}(x)p^{q}(x)dx-
\lambda_{2}\int_{0}^{\infty}ln_{q}^{2}(x)p^{q}(x)dx \\
\nonumber then \quad  p(x)=C[x^{q-1}+\frac{\lambda_{2}}{q-1}\left(x^{q-1}-1\right)^{2}]^{\frac{1}{1-q}} \qquad with \quad C=\left[\frac{\lambda_{0}(1-q)}{q}\right]^{\frac{1}{q-1}}
\end{eqnarray}
Where we have taken $\lambda_{1}=1$. Taking the limit $q\To 1$ inside the integrals we apply the first condition giving:
\begin{eqnarray}
\nonumber \lim_{q\To 1}\int_{0}^{\infty}p(x)dx=\int_{0}^{\infty}\lim_{q\To 1}p(x)dx=C\sqrt{\frac{\pi}{\lambda_{2}}}=1\quad \Longrightarrow \quad C=\sqrt{\frac{\lambda_{2}}{\pi}}
\end{eqnarray}
The third condition gives:
\begin{eqnarray}
\int_{0}^{\infty}\lim_{q\To 1}ln_{q}^{2}(x)p^{q}(x)dx=\frac{C\sqrt{\pi}}{2\lambda_{2}^{3/2}}=\frac{1}{2\lambda_{2}}=\sigma^{2}
\label{logNsigma}
\end{eqnarray}
The generalized Log-Normal distribution and the escort probabilities are the same:
\begin{eqnarray}
P_{esc}(x)=\left[\frac{\lambda_{2}}{\pi}\right]^{q/2} [x^{q-1}+\frac{\lambda_{2}}{q-1}\left(x^{q-1}-1\right)^{2}]^{\frac{q}{1-q}}\\
\nonumber \qquad \forall x \in I=(0,1),\quad q\in(1,2)
\end{eqnarray}
Taking the limit $q\To 1$ in this expression and using \ref{logNsigma} we get:
\begin{eqnarray}
\nonumber p(x)=\frac{1}{\sigma x\sqrt{2\pi}}\exp\left(-\frac{\ln^{2}x}{2\sigma^{2}}\right)
\end{eqnarray}
which is the Log-Normal distribution.

\subsection{Generalized Normal Distribution: $I=(-\infty,\infty)$} 
The constraints are: \\\\
$1.\;\int_{-\infty}^{\infty}p(x)dx=1 \qquad 2.\;\int_{-\infty}^{\infty}xp^{q}(x)dx=0\qquad 3.\;\int_{-\infty}^{\infty}x^{2}p^{q}(x)dx=\sigma^{2}$\\\\
We need to maximize:
\begin{eqnarray}
\nonumber S^{*}_{q}=S_{q}-\lambda_{0}\int_{-\infty}^{\infty}p(x)dx-\lambda_{1}\int_{-\infty}^{\infty}xp^{q}(x)dx-
\lambda_{2}\int_{-\infty}^{\infty}x^{2}p^{q}(x)dx \\
\nonumber then p(x)=C[1+\lambda_{1}x(q-1)+\lambda_{2}x^{2}(q-1)]  ^{-\frac{1}{q-1}}  \qquad with \quad C=\left[\frac{\lambda_{0}(1-q)}{q}\right]^{\frac{1}{q-1}}
\end{eqnarray}
If we take the limit $q\To 1$ for $p(x)$ we get:
\begin{eqnarray}
\nonumber \lim_{q\To 1}p(x)=Ce^{-\lambda_{1}x}e^{-\lambda_{2}x^{2}}
\end{eqnarray}
If we choose $\lambda_{1}=0$ we obtain the same results as for the Normal distribution, and the escort probabilities and the distribution ar the same:
\begin{eqnarray}
\nonumber C=\sqrt{\frac{\lambda_{2}}{\pi}}\\
\nonumber \lambda_{2}=\frac{1}{2\sigma^{2}}\\
P_{esc}(x)=\sqrt{\frac{\lambda_{2}}{\pi}}[1+\lambda_{2}x^{2}(q-1)]^{-\frac{1}{q-1}}
\end{eqnarray}
Is easy to verify that the limit of this expression is the Normal Distribution with null mean value.

\subsection{Generalized Laplace Distribution: $I=(-\infty,\infty)$} 
The constraints are: \\\\
$1.\;\int_{-\infty}^{\infty}p(x)dx=1\qquad 2.\;\int_{-\infty}^{\infty}\abs x p^{q}(x)dx=w$\\\\
We need to maximize:
\begin{eqnarray}
\nonumber S^{*}_{q}=S_{q}-\lambda_{0}\int_{-\infty}^{\infty}p(x)dx-\lambda_{1}\int_{-\infty}^{\infty}\abs x p^{q}(x)dx\\
\nonumber p(x)=C[1+\lambda_{1}\abs x(q-1)]^{\frac{1}{1-q}}\qquad with \quad C=\left[\frac{\lambda_{0}(1-q)}{q}\right]^{\frac{1}{q-1}}
\end{eqnarray}
Taking limit:
\begin{eqnarray}
\nonumber \lim_{q\To 1}\int_{-\infty}^{\infty}p(x)dx=\int_{-\infty}^{\infty}\lim_{q\To 1}p(x)dx=\int_{-\infty}^{\infty}Ce^{-\lambda_{1}\abs x}dx=\\
\nonumber =\frac{2C}{\lambda_{1}}=1\quad then\quad C=\frac{\lambda_{1}}{2}
\end{eqnarray}
This reproduces the known behavior of Laplace distribution. The escort probabilities and the distribution are the same:
\begin{eqnarray}
P_{esc}(x)=\left(\frac{\lambda_{1}}{2}\right)^{q}[1+\lambda_{1}\abs x(q-1)]^{\frac{q}{1-q}}
\end{eqnarray}
With the second condition we obtain:
\begin{eqnarray}
\nonumber \int_{-\infty}^{\infty}\abs x p^{q}(x)dx=\frac{2C}{\lambda_{1}^{2}}=\lambda_{1}=w
\end{eqnarray}

\bigskip

\noindent
\section*{Conclusions}
We have obtained mainly these results for the Tsallis entropy and we have shown that the use of Renyi entropy does not imply any new result because generalized versions are almost the same differing in a constant value. The generalized versions of the continuous statistical distributions depend on the parameter $q\in\mathbb{R}$ as the Tsallis entropy does. In the limit $q\rightarrow1$ we obtain the BGS entropy from the Tsallis generalization and so it happens with generalized versions of distributions where we can obtain the classical distributions using the same limit. The application of this procedure to other statistical distributions is straightforward.\\\\
The application of these generalized distributions to physical situations should be of interest and we hope they can be found very quickly. In particular, Brouers and Sotolongo \cite{Brouers} derived the dipolar relaxation function for a cluster model whose volume distribution was obtained from the generalized maximum Tsallis nonextensive entropy principle relating the power law exponents of the relaxation function to a global fractal parameter $\alpha$ and for large time to the entropy nonextensivity parameter $q$.

\end{document}